\documentclass[11pt]{article}
\usepackage[centertags]{amsmath}
\usepackage{amssymb}

\newtheorem{thm}{Theorem}[section]
\newtheorem{defn}[thm]{Definition}

\newtheorem{rem}[thm]{Remark}

\def\qed{\hfill {\vrule height5pt width5pt depth2pt}}

\begin{document}

\title{\bf Generalized Jarzynski's equality of inhomogeneous multidimensional diffusion processes}

\author{Hao Ge\footnote{LMAM, School of Mathematical Sciences,
                      Peking University, Beijing 100871, P.R. China; Email: edmund\_ge@tom.com}
        \and Da-Quan Jiang\footnote{LMAM, School of Mathematical Sciences,
                      Peking University, Beijing 100871, P.R. China; E-mail: jiangdq@math.pku.edu.cn}
       }
\maketitle{}

\begin{abstract}
Applying the well-known Feynman-Kac formula of inhomogeneous case,
an interesting and rigorous mathematical proof of generalized
Jarzynski's equality of inhomogeneous multidimensional diffusion
processes is presented, followed by an extension of the second law
of thermodynamics. Then, we explain its physical meaning and
applications, extending Hummer and Szabo's work ({\em Proc. Natl.
Acad. Sci. USA } {\bf 98}(7),  3658--3661 (2001)) and Hatano-Sasa
equality of steady state thermodynamics ({\em Phys. Rev. Lett.} {\bf
86}, 3463--3466 (2001)) to the general multidimensional case.
\begin{flushleft}
\end{flushleft}
\end{abstract}

\section{Introduction}

Thermodynamics of irreversible systems far from equilibrium has been
developed for more than thirty years since the original works by
Haken \cite{Hak77,Hak83} about laser and Prigogine, et al.
\cite{GlPr,NP} about oscillations of chemical reactions. A
nonequilibrium system can be regarded as an open system with
positive entropy production, which means exchange of substances and
energy with its environment.

At almost the same time, T.L. Hill, etc. \cite{Hi66,Hi77,Hi95,HC}
constructed a general mesoscopic model for the combination and
transformation of biochemical polymers in vivid metabolic systems
since 1966, which can be applied to explain the mechanism of muscle
contraction and active transports, such as the $Na$ and $K$ ions
actively transferring and penetrating through organic membranes in
the Hodgekin-Huxley model. These contributions have been developed
and summarized in \cite{Kei}.

One can use stationary homogeneous Markov chains and diffusion
processes as mathematical tools to model nonequilibrium steady
states and cycle fluxes. Mathematical theory of nonequilibrium
steady states has been developed for more than three decades since
Qians' original works \cite{QQ82,QQ85,QQG,QQQ84}. They derived the
formulae for entropy production rate and circulation distribution of
homogeneous Markov chains, Q-processes and diffusions. They
concluded that the chain or process is reversible if and only if its
entropy production rate vanishes, or iff there is no net cycle
fluxes. Here, we recommend a recent book \cite{JQQ} for the
systematic presentation of this theory.

In the past decade, a few relations that describe the statistical
dynamics of driven systems have been discovered, which are valid
even if the system is driven far from equilibrium. These include
Jarzynski's exciting nonequilibrium work relation
\cite{Jar1,Jar2,Jar3,Cro1,Cro2,Cro3}, which gives equilibrium
Helmholtz free energy differences in terms of nonequilibrium
measurements of the work required to switch from one ensemble to
another. This result has been applied to the mechanical extension of
single RNA molecules in the laboratory \cite{LDSTB}. Although the
concept of Helmholtz free energy fails in nonequilibrium steady
states (NESS), Hatano and Sasa \cite{HS} have generalized
Jarzynski's work to the NESS described by a simple one-dimensional
Langevin system, which is more relevant to motor proteins.

However, few rigorous mathematical results are derived since the
emergence of Jarzynski's equality. Applying the Feynman-Kac formula,
G. Hummer and  A. Szabo gave a quite brief proof of Jarzynski's
equality for inhomogeneous diffusive dynamics on a potential
\cite{HuS}, and after that, Hong Qian investigated a simple
two-state example of inhomogeneous Markov chains \cite{QH1}. But in
fact, their proofs are not mathematically rigorous, and they all
misused the Feynman-Kac formula of the inhomogeneous case
\cite[Theorem 5.7.6]{KS}, since it is quite different from the
Feynman-Kac formula of the homogeneous case \cite[Theorem 4.4.2]{KS}
and the former is actually more difficult to apply than the latter
one. Further explanation is included in Subsection 3.1. On the other
hand, the Jarzynski's equality is trivial in the homogeneous case
(see Remark \ref{rem-homo} below), which actually implies that
inhomogeneity is a necessity for Jarzynski's equality to make sense.

Recently enlightened by the work of Crooks \cite{Cro3}, we gave a
completely different and interesting rigorous derivation of  the
Jarzynski's equality in inhomogeneous Markov chains \cite{Ge3},
without applying the Feynman-Kac formula. Moreover, we investigated
the relationship between Jarzynski's equality and the statistical
physical property in the model of inhomogeneous Markov chains,
including reversibility and entropy production \cite{Ge1,Ge2}.

Nevertheless, it is not easy to extend the main idea of proof in
\cite{Ge3} to the case of inhomogeneous diffusion processes, because
by this way, one needs a very general version of the
Cameron-Martin-Girsanov formula similar to \cite[Thm. 6.4.2]{SV},
which is difficult to derive. On the other hand, physicists always
believe that inhomogeneous diffusion processes can be regarded as
the limit of inhomogeneous Markov chains, and in most of their
works, they actually only proved the corresponding results in the
case of inhomogeneous Markov chains rather than diffusion processes.
However, from the mathematical point of view, inhomogeneous
diffusion processes can only be regarded as the limit of
inhomogeneous Markov chains \emph{in distribution} rather than
\emph{in trajectories}. Hence, the Jarzynski's equality in
inhomogeneous diffusion processes can not be directly derived as the
limit in some sense of that in inhomogeneous Markov chains, and we
have to appeal to the Feynman-Kac formula of inhomogeneous case.

In this paper, applying the well-known Feynman-Kac formula of
inhomogeneous case, an interesting and rigorous mathematical proof
of generalized Jarzynski's equality in inhomogeneous
multidimensional diffusion processes is presented in Section 2,
followed by an extension of the second law of thermodynamics. It
should be mentioned that the method of proof in the present paper
can also be applied to derive the same Jarzynski's equality in
inhomogeneous Markov chains as \cite{Ge3}, or even possibly to
extend to general Markov processes. In Section 3, we explain its
physical meaning and applications, extending Hummer and Szabo's work
\cite{HuS} and Hatano-Sasa equality of steady state thermodynamics
\cite{HS} to the general multidimensional case.

In order to make the present paper accessible to a somewhat wider
audience, some reasonable sufficient conditions for Jarzynski's
equality of inhomogeneous diffusion processes are provided in Remark
2.7 below.

\section{Mathematical theory of generalized Jarzynski's equality}

\subsection{Basic property of inhomogeneous diffusion processes}

This subsection is about the construction of inhomogeneous diffusion
processes applying the fundamental solutions of partial differential
equations. The conditions given here are somewhat optimal, and the
readers who are not interested in technical details can directly
skip to the next subsection for the proof of generalized Jarzynski's
equality. We note here that most of these conditions including (A1),
(A2) and (A4) are satisfied when all the coefficients belong to the
smooth function set $C^{\infty}$ and all the derivatives are
uniformly bounded.

Denote $A_t(x)=(a_{ij}(t,x))_{d\times d}$ and
$\bar{b}_t(x)=(\bar{b}_i(t,x))_{d\times 1}$, where $a_{ij}(t,x)$ and
$\bar{b}_i(t,x)$ are functions defined on $[0,+\infty)\times
\mathbb{R}^d$.
Suppose that\\
(A1) $a_{ij}(t,x),\bar{b}_i(t,x)$ are uniformly bounded and
uniformly continuous with respect to
both $x$ and $t$, and  also satisfy a H\"{o}lder condition with respect to $x$;\\
(A2) $a_{ij}(t,x)$ satisfy a H\"{o}lder condition with respect to $t$;\\
(A3) $a_{ij}(t,x)$ satisfy the uniform ellipticity condition, i.e.
there exists $\gamma >0$, such that for
any $d$-dimensional real vector $\lambda=(\lambda_1,\lambda_2,\cdots,\lambda_d)$,\\
$$\sum_{i,j=1}^d a_{ij}(t,x)\lambda_i\lambda_j\geq \gamma\sum_{i=1}^d \lambda_i^2;$$
(A4) The derivatives $\frac{\partial a_{ij}}{\partial
x_i},\frac{\partial^2 a_{ij}}{\partial x_i\partial
x_j},\frac{\partial \bar{b}_i}{\partial x_i}$ exist, uniformly
bounded and satisfy a H\"{o}lder condition with respect to $x$.

For simplicity, let $b_t=(b_i(t,x))_{d\times 1}$ and
$b_i(t,x)=\bar{b}_i(t,x)-\frac{1}{2}\sum_{j=1}^d\frac{\partial
a_{ij}(t,x)}{\partial x_j}$.

Theorem \ref{cons1} below is rewritten from \cite[Vol II, Theorem
0.4, pp. 227]{Dy1} and \cite[Chap. 1, Sec. 6, Theorem 11, pp. 24;
Chap. 1, Sec. 8, Theorem 15]{F}, and Theorem \ref{cons2} is
rewritten from \cite[Vol I, remark of Theorem 5.11, pp. 167]{Dy1}
and \cite[Chap. 4]{Dy2}. One can also find the same results in
\cite[pp. 368-369]{KS} and \cite[Chapter 3]{SV}.

\begin{thm}\label{cons1}
If the coefficients satisfy conditions $(A1)$, $(A2)$, $(A3)$, then
the equation
\begin{equation}\label{eq1}
\frac{\partial u}{\partial s}+D_su=0,
\end{equation}
where $D_su(s,x)=\frac{1}{2}\sum_{i,j=1}^d
\frac{\partial}{\partial x_i} a_{ij}(s,x)\frac{\partial
u}{\partial x_j}+\sum_{i=1}^d b_i(s,x)\frac{\partial u}{\partial
x_i}=(\frac{1}{2}\nabla\cdot A(s,x)\nabla+b(s,x)\cdot\nabla)u$,
has a unique fundamental solution $p(s,t;x,y)$, satisfying:\\
$(B1)$: $p(s,t;x,y)>0$ for each $s$, $t$ and $x$, $y$;\\
$(B2)$: In addition, if coefficients $a_{ij}(t,x),\bar{b}_i(t,x)$
satisfy $(A4)$, then $p(s,t;x,y)$ satisfies the conjugate equation:

\begin{equation}
\frac{\partial u}{\partial t}=\bar{D}_t^*u,\label{eq2}
\end{equation}
where $\bar{D}_t^*u(t,y)=\sum_{i=1}^d \frac{\partial}{\partial
y_i}[\frac{1}{2} \sum_{j=1}^d a_{ij}(t,y)\frac{\partial
u(t,y)}{\partial  y_j}- b_i(t,y)u(t,y)]
=\nabla\cdot [\frac{1}{2}A(t,y)\nabla u(t,y)-b(t,y)u(t,y)]$;\\
$(B3)$: For any bounded continuous function $f(x)$, $u(s,t,x)=\int
p(s,t;x,y)f(y)dy$ satisfies (\ref{eq1}) and $\lim_{s\uparrow
t}u(s,t,x)=f(x)$, which is uniformly convergent in any bounded
domain of $\mathbb{R}^d$; $v(s,t,y)=\int p(s,t;x,y)f(x)dx$ satisfies
(\ref{eq2}), and $\lim_{t\downarrow s}v(s,t,y)=f(y)$, which is also
uniformly
convergent in any bounded domain of $\mathbb{R}^d$;\\
$(B4)$: The following inequalities are satisfied:
$$p(s,t;x,y)\leq M(t-s)^{-\frac{d}{2}}e^{-\frac{\alpha|y-x|^2}{t-s}};$$
$$\frac{\partial p(s,t;x,y)}{\partial x_i}\leq M(t-s)^{-\frac{d+1}{2}}e^{-\frac{\alpha|y-x|^2}{t-s}};$$
$$\frac{\partial^2 p(s,t;x,y)}{\partial x_i\partial x_j}\leq M(t-s)^{-\frac{d}{2}-1}e^{-\frac{\alpha|y-x|^2}{t-s}};$$
$$\frac{\partial p(s,t;x,y)}{\partial t}\leq M(t-s)^{-\frac{d}{2}-1}e^{-\frac{\alpha|y-x|^2}{t-s}};$$
$$p(s,t;x,y)\geq M_1(t-s)^{-\frac{d}{2}}e^{-\frac{\alpha_1|y-x|^2}{t-s}}-M_2(t-s)^{-\frac{d}{2}+\lambda}e^{-\frac{\alpha_2|y-x|^2}{t-s}},$$
where $M,M_1,M_2$ and $\alpha,\alpha_1,\alpha_2$ are all positive constants.\\
$(B5)$: If $f(x)$ is a bounded function with second-order continuous
derivatives, which satisfy a H\"{o}lder condition, then
$u(s,t,x)=\int p(s,t;x,y)f(y)dy$ satisfies
$$\lim_{s \uparrow t}\frac{\partial u}{\partial x_i}=\frac{\partial f}{\partial x_i},~\lim_{s \uparrow t}\frac{\partial^2 u}{\partial x_i\partial x_j}=\frac{\partial^2 f}{\partial x_i\partial x_j};$$
while $v(s,t,y)=\int p(s,t;x,y)f(x)dx$ satisfies
$$\lim_{t \downarrow s}\frac{\partial v}{\partial y_i}=\frac{\partial f}{\partial y_i},~\lim_{t \downarrow s}\frac{\partial^2 v}{\partial y_i\partial y_j}=\frac{\partial^2 f}{\partial y_i\partial y_j}.$$
\end{thm}

\begin{rem}
{\rm
Most of the following definitions of physical quantities make sense
due to the basic inequalities in $(B4)$ together with $(B3)$ and
$(B5)$.}
\end{rem}

\begin{thm}\label{cons2}
There exists a unique inhomogeneous diffusion process $X=\{X_t:
t\geq 0\}$ on $\mathbb{R}^d$, whose transition probability density
is $\{p(s,t;x,y)\}$. Moreover, $X$ is a strong Markov process. We
call $X$ the {\bf diffusion process with infinitesimal generator $D$
}.
\end{thm}

Equation (\ref{eq1}) is called the  backward Kolmogorov equation of
$X$, while equation (\ref{eq2}) is called the forward Kolmogorov
 equation of $X$.

Denote the initial distribution density of $X$ as
$\{\rho_0(x)>0:x\in \mathbb{R}^d\}$, which is at least twice
differentiable, then $\rho_t(x)=\int \rho_0(y)p(0,t;y,x)dy$ is the
density function of $X_t$, simply denoted as $\rho_t$. Thus from
(B3), $\rho_t(x)$ satisfies (\ref{eq2}), which is called the
Fokker-Planck equation.

Indeed, due to the condition $(A3)$, there exists a nonsingular
$d\times d$ matrix $\Gamma_t(x)=(\Gamma_{ij}(t,x))_{d\times d}$ such
that $A_t(x)=\Gamma_t(x)\Gamma^T_t(x)$, where $\Gamma^T_t(x)$ is the
transpose matrix of $\Gamma_t(x)$. The inhomogeneous
multidimensional diffusion process $\{X_t: t\geq 0\}$ can be considered as the unique
solution of the stochastic differential equation
$$dX_t=\bar{b}_t(X_t)dt+\Gamma_t(X_t)dW_t,$$
where $\{W_t\}_{t\geq 0}$ is a $d$-dimensional Wiener process.

\subsection{Rigorous proof of generalized Jarzynski's equality}

Fix the time interval as $[0,T]$. In order to make the quantities in
Jarzynski's equality below mathematically well-defined, we have to
make another basic assumption:

(A5) The elliptic equation $\bar D^*_tf(x)=0$ has a unique strong
$L^1$ solution $\pi_t=\{\pi_t(x): x \in \mathbb{R}^d\}$ such that
$\int_{\mathbb{R}^d} \pi_t(x)dx\equiv 1$, recalling
$\bar{D}_t^*\pi_t(x)=\sum_{i=1}^d \frac{\partial}{\partial
x_i}[\frac{1}{2} \sum_{j=1}^d a_{ij}(t,x)\frac{\partial
\pi_t(x)}{\partial  x_j}- b_i(t,x)\pi_t(x)] =\nabla\cdot
[\frac{1}{2}A(t,x)\nabla \pi_t(x)-b(t,x)\pi_t(x)]$. Moreover,
$\pi_t(x)$ is continuously differentiable and uniformly bounded with
respect to parameter $t\in [0,T]$; $\frac{\partial
[\pi_t(x)]}{\partial t}$ is uniformly bounded too for $t\in [0,T]$.
In addition, suppose $\pi_t(x)>0$, $\forall x\in \mathbb{R}^d$,
$t\geq 0$.

\begin{rem} \label{rem-inv-dist}
{\rm Consider a homogenous diffusion process with infinitesimal
generator $D=\frac{1}{2}\nabla\cdot A(x)\nabla+b(x)\cdot\nabla$. In
the uniformly elliptic case, if $\bar D^*\pi(x)=0$ has a positive
strong $L^1$ solution such that $\int_{R^d}\pi(x)dx=1$, it is
unique. To have a solution one has to impose a sufficiently strong
inward drift at infinity, or equivalently suppose that the
homogeneous diffusion process is positive recurrent \cite[Chap
IV]{Has80}.

Physicists are interested in some reasonable sufficient conditions.
A particular example is the equilibrium case discussed in the next
section. More generally, it is sufficient to suppose that the
diffusion coefficient $A(x)$ is bounded and uniformly elliptic,
$b(x)$ is bounded smooth, and for large $x$, it is the minus
gradient of a confining potential $U(x)$, which is satisfied for
many physical problems. The proof can be based on the Lyapunov
function criteria for asymptotic stability of stochastic dynamic
systems (\cite[Theorem 11.9.1]{LM} and \cite[Theorem
III.5.1]{Has80}):

If there exists a smooth function $V(x)$ with the properties
$$V(x)\geq 0$$
and
$$\lim_{R\rightarrow\infty} \sup_{|x|>R}DV(x)=-\infty,$$
then there exists a stationary distribution for the diffusion
process.

Therefore, if the potential $U(x)$ satisfies that
\begin{eqnarray}\label{Sufficond}
&&\sup_{|x|>R}\left[\frac{1}{2}\nabla\cdot A(x)\nabla
U(x)+b(x)\cdot\nabla
U(x)\right]\nonumber\\
&&=\sup_{|x|>R}\left[\frac{1}{2}\nabla\cdot A(x)\nabla
U(x)-\parallel
\nabla U(x)\parallel_2^2\right]\nonumber\\
&&\rightarrow -\infty {\rm ~as~}R\rightarrow\infty,
\end{eqnarray}
where $\parallel \vec v\parallel_2^2=\sum_{i=1}^d v_i^2$, then $\bar
D^*\pi(x)=0$ has a unique positive strong $L^1$ solution such that
$\int_{R^d}\pi(x)dx=1$. For example, the sufficient condition
(\ref{Sufficond}) is satisfied when the potential $U(x)$ is a
polynomial with even highest order.

Another nontrivial example is the multidimensional Ornstein-Ulenbeck
process with drift coefficients $Bx=(b_{ij})_{d\times d}\cdot x$ and
diffusion coefficients $\sigma=\{\sigma_{ij}\}_{d\times r}$. Its
corresponding Fokker-Planck equation is
$$\frac{\partial u}{\partial t}=\frac{1}{2}\sum_{i,j=1}^d a_{ij}\frac{\partial^2 u}{\partial x_i\partial x_j}-\sum_{i=1}^d\frac{\partial}{\partial x_i}[b_i(x)u],$$
where $b_i(x)=\sum_{j=1}^d b_{ij}x_j$ and $a_{ij}=\sum_{k=1}^r
\sigma_{ik}\sigma_{jk}$. It is well known that its unique stationary
distribution is a multidimensional normal distribution with mean
zero and variance $\Sigma=\int_0^{+\infty} e^{Bs}Ae^{B^Ts}ds$,
provided that the matrix $A=(a_{ij})$ is nonsingular and the real
parts of all eigenvalues of $B=(b_{ij})_{d\times d}$ are negative.

According to \cite[Example 11.9.2]{LM}, the Ornstein-Ulenbeck
semigroup determined by the above Fokker-Planck equation is
asymptotically stable with limiting density of $N(0,\Sigma)$.
Moreover, according to \cite[Theorem 3.3.7]{JQQ}, the stationary
multidimensional Ornstein-Ulenbeck process is reversible (or say, in
equilibrium state) if and only if the force $F=2A^{-1}Bx$ is the
gradient of a potential $U(x)$ satisfying $\int e^{U(x)}dx=1$, of
iff the coefficient $A$ and $B$ satisfy the symmetry condition
$A^{-1}B=(A^{-1}B)^T$ \cite{QH2}. Therefore, nonequilibrium
Ornstein-Ulenbeck processes can only exist in the multidimensional
($d\geq 2$) case.

One could also require that the diffusion process is confined to
some compact set or manifold (e.g. torus), and the same arguments
below can also be applied to prove the Jarzynski equality in those
cases without any difficulties.}
\end{rem}

$\pi_t$ can be called  the {\bf quasi-invariant
distribution}\footnote{The notion ``quasi-invariant distribution''
means that if one takes $A(t,x)$ and $\bar{b}(t,x)$ as the diffusion
coefficient and drift coefficient of a homogeneous diffusion
process respectively, $\pi_t(x)$ is just its invariant
distribution.} at time $t$.

Fix the initial distribution density $\rho_0=\pi_0$, which is one of the key points in Jarzynski's
equality. Given an arbitrary absolutely continuous function $F(t)$,
denote
\begin{equation}
 H(t,x)=F(t)-\beta^{-1}\log{\pi_t(x)},~\forall x\in \mathbb{R}^d.
\end{equation}
Then
$$\pi_t(x)=\frac{e^{-\beta H(t,x)}}{\int_{\mathbb{R}^d} e^{-\beta H(t,x)}dx},$$
and
$$F(t)=-\beta^{-1}\ln\int_{\mathbb{R}^d} e^{-\beta H(t,x)}dx,$$
where $\beta>0$ is a constant. Let
\begin{eqnarray}
W(\omega)&=&\int_0^T \frac{\partial H}{\partial s}(s,X_s)ds\nonumber\\
&=&\int_0^T \frac{\partial [F(s)-\frac{1}{\beta}\log\pi_s(X_s)]}{\partial s}ds\nonumber\\
&=&\triangle F-\int_0^T \frac{\partial
[\frac{1}{\beta}\log\pi_s](X_s)}{\partial s}ds,
\end{eqnarray}
where $\triangle F=F(T)-F(0)$. Write
\begin{eqnarray}
\triangle H(\omega)&=&H(T,X_T)-H(0,X_0),
\end{eqnarray}
and
\begin{eqnarray}\label{Fir-Ther}
Q(\omega)&=&\triangle H(\omega)-W(\omega).
\end{eqnarray}

The following theorem is the basis of our proof, which is an
application of the famous Feynman-Kac formula in inhomogeneous case
\cite[Theorem 5.7.6]{KS}.

\begin{thm}\label{Fey-Kac}
Under the preceding assumptions, let $W_d=W-\triangle F=-\int_0^T
\frac{\partial [\frac{1}{\beta}\log\pi_s]}{\partial s}(X_s)ds$, and
denote
\begin{equation}\label{Def-v}
v(t,x)=E_{t,x}\exp\left[\int_t^T \frac{\partial
[\log\pi_s]}{\partial s}(X_s)ds\right],
\end{equation}
where $E_{t,x}$ means the expectation is taken conditioned on the
event $\{X_t=x\}$, then $v(t,x)$ satisfies the Cauchy problem
\begin{equation}\label{Cauchy}
\left \{\begin{array}{l}\frac{\partial v}{\partial
t}(t,x)=-\frac{\partial
[\log\pi_t(x)]}{\partial t}v(t,x)-D_tv(t,x),\\
v(T,x)=1,\end{array}\right.
\end{equation}
recalling that $D_tv(t,x)=(\frac{1}{2}\nabla\cdot
A(t,x)\nabla+b(t,x)\cdot\nabla)v(t,x)$. Moreover, such a solution is
unique.
\end{thm}
\emph{Proof: } We only need to check the condition of Theorem 5.7.6
in \cite{KS}.

First, according to \cite[Vol II, Theorem 0.4, pp. 227]{Dy1}, the
Cauchy problem (\ref{Cauchy}) has a solution $v(t,x)$ which is
continuous and satisfies the exponential growth condition
$$\max_{0\leq t\leq T}|v(t,x)|\leq Me^{\mu||x||^2},~x\in \mathbb{R}^d,$$
for some constants $M>0$ and $\mu>0$.

Moreover, due to the condition $(A1)$, the coefficients
$a_{ij}(t,x)$ and $\bar{b}_i(t,x)$ are uniformly bounded with
respect to $t$ and $x$.

Finally, applying \cite[Theorem 5.7.6 and Problem 5.7.7]{KS},
together with Theorem \ref{cons1} and Theorem \ref{cons2}, we get
the desired result. \qed

Now, it is time to derive the generalized Jarzynski's equality of
multidimensional diffusion processes.
\begin{thm}\label{Jar's eq} Under the preceding assumptions, suppose that

(i) $\int \pi_t(x)v(t,x)dx<+\infty$ for $t\in [0,T]$, and  $\int
\frac{\partial[\pi_t(x)v(t,x)]}{\partial t}dx$ are uniformly
convergent for $t\in [0,T]$; and

(ii) for each $i$ and $j$,
\begin{eqnarray}\label{condition}
&&\lim_{x\rightarrow\infty}\pi_t(x)b_i(t,x)v(t,x)=0,\nonumber\\
&&\lim_{x\rightarrow\infty}\pi_t(x)a_{ij}(t,x)\frac{\partial v(t,x)}{\partial x_j}=0,\nonumber\\
&&\lim_{x\rightarrow\infty}\frac{\partial \pi_t(x)}{\partial
x_i}a_{ij}(t,x)v(t,x)=0;\nonumber
\end{eqnarray}
then we have
\begin{equation}\label{Jar's equation}
E^{P_{[0,T]}}[e^{-\beta W_d}]=1,
\end{equation}
i.e.
$$E^{P_{[0,T]}}[e^{-\beta W}]=e^{-\beta \triangle F}.$$
\end{thm}
{\em Proof:} Let $g(t)=\int \pi_t(x)v(t,x)dx$, our aim is to show
that $g(t)\equiv 1$, $\forall t\in [0,T]$.

Firstly, due to the assumption (i), it holds that
\begin{equation}
\frac{dg(t)}{dt}=\int \frac{\partial[\pi_t(x)v(t,x)]}{\partial
t}dx=\int \left [\frac{\partial \pi_t(x)}{\partial
t}v(t,x)+\pi_t(x)\frac{\partial v}{\partial t}(t,x)\right ]dx,\nonumber
\end{equation}
according to \cite[Vol. III, Section 21.3, Theorem 4,  pp.395]{ZZS}.
And from Theorem \ref{Fey-Kac}, follows that
\begin{eqnarray}
&&\frac{\partial\pi_t(x)}{\partial t}v(t,x)+\pi_t(x)\frac{\partial
v}{\partial
t}(t,x)dx\nonumber\\
&=&\frac{\partial\pi_t(x)}{\partial
t}v(t,x)+\pi_t(x)\left [-\frac{\partial [\log\pi_t(x)]}{\partial
t}v(t,x)-D_tv(t,x)\right ]\nonumber\\
&=&-\pi_t(x)D_tv(t,x).\nonumber
\end{eqnarray}
Then, with the assumption (ii), integrating by parts, one has
\begin{equation}
\frac{dg(t)}{dt}=\int [-\pi_t(x)D_tv(t,x)]dx=\int
[-v(t,x)\bar{D}_t^*\pi_t(x)]dx=0,~\forall t\geq 0,\nonumber
\end{equation}
which together with the fact that $g(T)=1$ implies $g(t)\equiv 1$.
Therefore, $g(0)=E^{P_{[0,T]}}[e^{-\beta W_d}]=1$. \qed

\begin{rem}{\bf Reasonable sufficient conditions}  \\
{\rm Some physicists may think that the conditions given in this
paper are more general than ever needed, and they would be satisfied
with reasonable sufficient conditions for the validity of the
Jarzynski equality. As we have mentioned at the beginning of this
section, it is sufficient to assume (a) all the coefficients are
bounded, at least twice continuously differentiable and all the
derivatives are uniformly bounded too; (b) the diffusion
coefficients $A_t(x)$ are uniformly elliptic (see (A3)); and (c) the
existence of quasi-invariant distributions (see (A5) and Remark
\ref{rem-inv-dist}). However, the technical requirements (i) and
(ii) in Theorem \ref{Jar's eq} are still not easy to be verified,
because we do not have exact estimation on the quantity $v(t,x)$.
But we believe that the conditions (a), (b) and (c) should be able
to guarantee the technical requirements (i) and (ii). Especially, in
the equilibrium case (see Subsection 3.1), it would be sufficient to
suppose that $H(t,x)\rightarrow \infty$ and $|\frac{\partial
H(t,x)}{\partial x_i}|\rightarrow \infty$ as $x\rightarrow\infty$,
and $x_i\cdot\frac{\partial H(t,x)}{\partial x_i}>0$ for all $i$ and
sufficiently large $x$.

In addition, one can explicitly calculate the quantity $v(t,x)$ in
some very special case. For example, consider the one-dimensional
diffusion dynamics
$$dX_t=-(X_t-t)dt+\sigma dW_t$$
on a moving potential $U(t,x)=\frac{1}{2}(x-t)^2$, which is
investigated by van Zon and Cohen \cite{vZonC03a, vZonC03b, vZonC04}
and mentioned in \cite{BJMS}. Its quasi-invariant distribution
$\pi_t(x)=\frac{1}{\sqrt{\pi} \sigma}e^{-\frac{(x-t)^2}{\sigma^2}}$.
Consequently, $H(t,x)=\frac{(x-t)^2}{\sigma^2}$, the Helmholtz free
energy $F(t)\equiv -\log\sqrt{\pi} \sigma$, and
\begin{eqnarray*}
 W_d&=&W=\int_0^T \frac{\partial H}{\partial s}(s,X_s)ds \\
    &=&-\frac{2}{\sigma^2}\int_0^T (X_s-s)ds.
\end{eqnarray*}
The unique solution of this stochastic differential equation can be
expressed as
$$X_s=e^{t-s}X_t+\left(s+e^{t-s}-te^{t-s}-1\right)+\sigma e^{-s}\int_t^s e^udW_u,~\forall ~s\geq t\geq 0.$$
Let
\begin{eqnarray*}
Y_t^T&=&-\frac{2}{\sigma^2}\int_t^T (X_s-s)ds \\
     &=&-\frac{2}{\sigma^2}\int_t^T\left(e^{t-s}X_t+e^{t-s}-te^{t-s}-1
        +\sigma e^{-s}\int_t^s e^udW_u\right)ds.
\end{eqnarray*}
Because the process $\{X_t:t\geq 0\}$ is Gaussian, one only needs to
calculate the expectation and variance of the quantity $Y_t^T$:
\begin{eqnarray*}
E_{t,x}Y_t^T&=&-\frac{2}{\sigma^2}\int_t^T\left(e^{t-s}x+e^{t-s}-te^{t-s}-1\right)ds\\
            &=&-\frac{2}{\sigma^2}\left[(x+1-t)\left(1-e^{t-T}\right)-T+t\right],
\end{eqnarray*}
and
\begin{eqnarray}
Var_{t,x}Y_t^T&=&Var_{t,x}\left[\frac{2}{\sigma}\int_t^Te^{-s}\int_t^s
e^udW_uds\right]\nonumber\\
&=&Var_{t,x}\left[\frac{2}{\sigma}\int_t^Te^udW_u\int_u^Te^{-s}
ds\right]\nonumber\\
&=&\frac{4}{\sigma^2}\int_t^T\left(\int_u^Te^{u-s}ds\right)^2du\nonumber\\
&=&\frac{4}{\sigma^2}\left[T-t-2\left(1-e^{t-T}\right)+\frac{1}{2}\left(1-e^{2t-2T}\right)\right]\nonumber.
\end{eqnarray}
Therefore,
\begin{eqnarray*}
 v(t,x)&=&E_{t,x}e^{-Y_t^T}
         =\exp\left[-E_{t,x}Y_t^T+\frac{Var_{t,x}Y_t^T}{2}\right] \\
       &=&\exp \left\{\frac{2}{\sigma^2}\left[x\left(1-e^{t-T}\right)-\frac{1}{2}-t
           +(t+1)e^{t-T}-\frac{1}{2}e^{2t-2T} \right]\right\}.
\end{eqnarray*}
Finally, it is easy to check that the process $\{X_t:t\geq 0\}$
satisfies the requirements (i) and (ii) in Theorem \ref{Jar's eq}
and the quantity
\begin{eqnarray*}
 g(t)&=&\int \pi_t(x)v(t,x)dx \\
     &=&\frac{1}{\sqrt{\pi}\sigma}\int \exp\left[-\frac{\left(x-t-1+e^{t-T}\right)^2}{\sigma^2}\right]dx\equiv 1.
\end{eqnarray*}
      }
\end{rem}




\begin{defn}
We call the inhomogeneous multidimensional diffusion process
$X=\{X(t)\}_{0\leq t\leq T}$ is {\bf unperturbed}, if $\pi_t$,
$0\leq t\leq T$, are all the same.
\end{defn}

The following is an extension of the second law of thermodynamics.
\begin{thm}\label{extend-Ther}
$E^{P_{[0,T]}}W_d\geq 0$, i.e.
\begin{equation}
 E^{P_{[0,T]}}W\geq \triangle F.
\end{equation}
Moreover, $E^{P_{[0,T]}}W=\triangle F$ if and only if
\begin{equation}
\int_0^T \frac{\partial \log\pi_s}{\partial
s}(X_s(\omega))ds=0,~P_{[0,T]}-a.s.\nonumber
\end{equation}
If the inhomogeneous multidimensional diffusion process
$X=\{X(t)\}_{0\leq t\leq T}$ is unperturbed, then
$W(\omega)=\triangle F$ for each trajectory $\omega$.
\end{thm}
{\em Proof:} By the Jazynski's equality (\ref{Jar's equation}) and
Jensen's inequality,
$$e^{-\beta\triangle F}=E^{P_{[0,T]}}[e^{-\beta W}]\geq e^{-\beta E^{P_{[0,T]}}W},$$
i.e.
$$E^{P_{[0,T]}}W\geq \triangle F,$$
and the equality holds if and only if $W(\omega)=\triangle F~a.s.$,
i.e.
\begin{equation}
\int_0^T \frac{\partial \log\pi_s}{\partial
s}(X_s(\omega))ds=0,~P_{[0,T]}-a.s.\nonumber
\end{equation}
Furthermore, if the inhomogeneous multidimensional diffusion process
$X=\{X(t)\}_{0\leq t\leq T}$ is unperturbed, then $\frac{\partial
\log\pi_s}{\partial s}(x)\equiv 0$, and consequently
$W(\omega)=\triangle F$ for each trajectory $\omega$.\qed

\begin{rem}\label{rem-homo}
{\rm In the homogeneous(steady state) case, the diffusion process is
obviously unperturbed, which yields $W_d\equiv 0$. So the theorems
above become trivial.}
\end{rem}

Recently, M. Baiesi et al. \cite{BJMS} have proved an exact
fluctuation theorem for the dissipative work $W_d$, i.e.
\begin{equation}\label{FT}
\frac{P_{\pi_0}(W_d=x)}{P_{\pi_0}(W_d=-x)}=e^{\beta x},
\end{equation}
for each $x$, under some condition \cite[Eq. 14]{BJMS}, which can
also give rise to the generalized Jarzynski's equality $Ee^{-\beta
W_d}=1$. They also pointed out that the diffusive dynamics on an
asymmetric potential may not agree with this exact fluctuation
theorem \cite[Fig. 2]{BJMS}, while the generalized Jarzynski's
equality (Theorem \ref{Jar's eq}) still holds. In addition, if the
time-averaged dissipative work $\frac{W_d}{T}$ has a large deviation
property with rate function $I(x)$, then (\ref{FT}) leads to the
Gallavotti-Cohen type fluctuation theorem $I(x)=I(-x)-x$.

\section{Physical meaning and applications}

\subsection{Generalization of Hummer and Szabo's work}

In Jarzynski's original work \cite{Jar1,Jar2,Jar3,Jar4} and Crooks'
recent work \cite{Cro1,Cro2,Cro3}, they derived the Jarzynski's
equality through standard derivation of statistical physics for
simple stochastic models including inhomogeneous denumerable Markov
chains.

Afterwards, in Hummer and Szabo's paper \cite{HuS}, it is shown how
equilibrium free energy profiles can be extracted rigorously from
repeated nonequilibrium force measurements on the basis of an
extension of Jarzynski's remarkable identity between free energies
and the irreversible work. But they misused the Feynman-Kac formula
\cite[Theorem 4.4.2, Theorem 5.7.6]{KS}. This well-known formula
provides a stochastic representation for the solution $v(t,x)$ of
the concerned parabolic equation through the conditional expectation
of the path integration of a specific stochastic process
$\xi=\{\xi_t: t\geq 0\}$, and $v(t,x)$ can be regarded intuitively
as an expectation with respect to the ``weighted'' phase space
distribution (see (\ref{Def-v})). In the inhomogeneous case, this
expectation is taken conditioned on the event $\{\xi_t=x\}$ at time
$t$, while in the homogeneous case, $x$ is taken to be the starting
point of the trajectory of $\xi$, i.e. the expectation in the
representation is taken conditioned on $\{\xi_0=x\}$ at the initial
time. More important, the quantity $v(t,x)$ in the Feynman-Kac
formula actually should be defined by holding the final time $T$
fixed and treating the time $t$ as a variable, and in the
inhomogeneous situation, the time interval for the path integration
can not be shifted from $[t,T]$ to $[0,T-t]$ while it can in the
homogeneous case, which implies that it is impossible to directly
integrate the quantity $v(t,x)$ with respect to $x$ applying the
initial equilibrium distribution to derive the generalized
Jarzynski's equality. Therefore, from the mathematical point of
view, the right side of \cite[Eq. 4]{HuS} and the quantity $g(z,t)$
defined in \cite[Eq. 14]{Jar2} can not satisfy the Feymann-Kac
formula in the inhomogeneous case. On the other hand, although the
Kolmogorov forward equation is more familiar and intuitively natural
for physicists, the path integration of the specific diffusion
process (i.e. the quantity $v(t,x)$) can only satisfy the elliptic
equation similar to the Kolmogorov backward equation according to
the standard Feynman-Kac formula \cite[Theorem 5.7.6]{KS}, which is
just another reason why Hummer and Szabo's derivation \cite{HuS} is
flawed.

What they considered is the diffusive dynamics on a potential
$V(t,x)$ satisfying $\int_{\mathbb{R}^d}V(t,x)dx<\infty$, whose time
evolution is governed by the differential operator
$L_t=\frac{1}{2}\nabla\cdot A(t,x)\nabla+b(t,x)\cdot\nabla$, in
which $A(t,x)=2D$ is the diffusion coefficient and $b(t,x)=D\nabla
V(t,x)$ is the drift coefficient. Now, we can restate the results of
Hummer and Szabo's work in the case of general multidimensional
diffusion processes, applying the mathematical theory in the
previous section. The statements below could be anticipated starting
from \cite{HuS}, but they are here rigorously derived.

It is important to point out that in the homogeneous diffusion case,
the force $2A^{-1}(x)b(x)$ has a potential $V(x)$ if and only if the
steady state is an equilibrium state \cite[Theorem 3.3.7]{JQQ}. In
this case, the invariant distribution $\pi(x)$ can be expressed as
the Boltzmann distribution

$$\pi(x)=\frac{e^{-V(x)}}{\int_{\mathbb{R}^d} e^{-V(x)} dx};$$
Moreover, the stationary diffusion process with initial distribution density $\pi(x)$ is in detailed balance.

Suppose that for each $t\in [0,T]$, the force $2A^{-1}(t,x)b(t,x)$
has a potential $-\beta H(t,x)$, then the quasi-invariant
distribution

$$\pi(t,x)=\frac{e^{-\beta H(t,x)}}{\int_{\mathbb{R}^d} e^{-\beta H(t,x)} dx},$$
and
$$F(t)=-\beta^{-1}\ln\int_{\mathbb{R}^d} e^{-\beta H(t,x)}dx$$ is the Helmholtz free
energy of this diffusion process at time $t$.

Therefore, $W(\omega)=\int_0^T \frac{\partial H}{\partial
s}(s,X_s)ds$ defined in the previous section is just the  external
work done on the system, $Q(\omega)$ in (\ref{Fir-Ther}) is regarded
as the total heat exchanged with the reservoir, and (\ref{Fir-Ther})
is actually the extension of the first law of thermodynamics.

The reversible work, $W_r=\triangle F=F(T)-F(0)$, is the free energy
difference between two equilibrium ensembles. And the dissipative
work $W_d=W-W_r$, is defined as the difference between the actual
work and the reversible work.

By  Theorem \ref{Jar's eq} and \ref{extend-Ther}, we get
\begin{thm}
Under the condition of Theorem \ref{Jar's eq}, the well known
Jarzynski's equality becomes
$$E^{P_{[0,T]}}[e^{-\beta W_d}]=1,$$ i.e.
\begin{equation}
 E^{P_{[0,T]}}[e^{-\beta W}]=e^{-\beta\triangle F}.
\end{equation}
And $E^{P_{[0,T]}}W_d\geq 0$, i.e. $E^{P_{[0,T]}}W\geq \triangle F$,
which is an extension of the second law of thermodynamics. Moreover,
$E^{P_{[0,T]}}W=\triangle F$ if and only if
\begin{equation}
\int_0^T \frac{\partial \log\pi_s}{\partial
s}(X_s(\omega))ds=0,~P_{[0,T]}-a.s.\nonumber
\end{equation}
If the inhomogeneous multidimensional diffusion process
$X=\{X(t)\}_{0\leq t\leq T}$ is unperturbed, then
$W(\omega)=\triangle F$ for each trajectory $\omega$.
\end{thm}

\begin{rem}
{\rm Although the existence of a potential for the force
$2A^{-1}b(t,x)$ is not a necessity in our proof of Section 2, it is
essential for the concept of free energy in physics, because free
energy can only be defined for the equilibrium states.}
\end{rem}

\subsection{Generalization of Hatano and Sasa's work}

As we have mentioned in the introduction of the present paper,
Hatano and Sasa only derived their result in the case of
inhomogeneous Markov chains rather than diffusion processes, and
they regarded the corresponding result in the case of inhomogeneous
diffusion processes be the direct limit of that in inhomogeneous
Markov chain case. From the mathematical point of view, diffusion
processes can be regarded as the limit of inhomogeneous Markov
chains, but only in the sense of distribution rather than
trajectories.

Using the phenomenological framework of steady-state
thermodynamics constructed by Oono and Paniconi \cite{OP}, they show
that an extended form of the second law holds for transitions
between steady states, relating the Shannon entropy (also accepted as
the common definition of Gibbs entropy) difference to the
excess heat produced in an infinitely slow operation. A generalized
version of the Jarzynski work relation plays an important role in
their theory \cite{HS}.

In their work, they studied a simple one-dimensional stochastic
model of Langevin dynamics describing nonequilibirum steady states
with drift coefficient $\frac{1}{\gamma}(-\frac{\partial
U(x;\alpha)}{\partial x}+f)$ and diffusion coefficient
$2k_BT=\frac{2}{\beta}$. What they are concerned with is to
establish the connection between the phenomena displayed by
nonequilibrium steady states and thermodynamic laws. Three kinds of
heats are defined: the total heat $Q_{tot}$, the housekeeping heat
$Q_{hk}$ and the excess heat $Q_{ex}$, satisfying
$Q_{tot}=Q_{hk}+Q_{ex}$. By convention, they take the sign of heat
to be positive when it flows from the system to the heat bath.

In the case of inhomogeneous multidimensional diffusion processes,
the housekeeping heat
\begin{equation}
Q_{hk}(\omega)=\frac{1}{\beta}\int_0^T [2A^{-1}b(t,X_t)-\nabla \log
\pi_t(X_t)]dX_t,
\end{equation}
where $dX_t$ is of the Stratonovich type. A simple example of this
interpretation of the heat was implied in Sekimoto's work \cite{Sek}
and explicitly defined in Hatano and Sasa's work \cite{HS}.

It has been proved that for equilibrium systems,
$2A^{-1}b(t,x)=\nabla \log \pi_t(x)$ \cite[Theorem 3.3.7]{JQQ},
hence $Q_{ex}$ reduces to the total heat $Q_{tot}$.

In the case of inhomogeneous multidimensional diffusion processes,
the total heat is defined as
\begin{equation}
Q_{tot}(\omega)=\frac{1}{\beta}\int_0^T
 (2A^{-1}b(t,X_t))dX_t.
\end{equation}
Since
$$\triangle H(\omega)=\int_0^T
\frac{\partial H}{\partial t}(t,X_t)dt+\int_0^T \nabla
H(t,X_t)dX_t$$ and
$$\nabla
H(t,x)=-\frac{1}{\beta}\nabla \log\pi_t(x),$$ we find that the
excess heat defined in Hatano and Sasa's paper is just
\begin{eqnarray}
Q_{ex}(\omega)&=&Q_{tot}(\omega)-Q_{hk}(\omega)\nonumber\\
&=&\int_0^T \frac{1}{\beta}\nabla \log\pi_t(X_t)dX_t\nonumber\\
&=&-\int_0^T \nabla H(t,X_t)dX_t\nonumber\\
&=&-\triangle H(\omega)+\int_0^T
\frac{\partial H}{\partial t}(t,X_t)dt\nonumber\\
&=&-Q(\omega),
\end{eqnarray}
where $Q(\omega)$ is defined in (\ref{Fir-Ther}).

Denote $\phi(t,x)=-\log\pi(t,x)$ to be
the Gibbs entropy (also called Gibbs free energy) of state $x$ at
time $t$, then
$$\triangle \phi=\beta(\triangle H-\triangle F).$$
Therefore,
\begin{equation} W_d=W-\triangle F=(\triangle
H-Q)-\triangle F=Q_{ex}+\frac{\triangle \phi}{\beta}.
\end{equation}
So we can get the generalized Jayzynski's equality of nonequilibrium
steady states.
\begin{thm}Under the condition of Theorem \ref{Jar's eq},
\begin{equation}
E^{P_{[0,T]}}[e^{-\beta Q_{ex}-\triangle\phi}]=1.
\end{equation}
\end{thm}

Let $S(t)=\langle \phi(t)\rangle=-\int_{\mathbb{R}^d}\pi(t,x)\log{\pi(t,x)}dx$ be
the Gibbs entropy (Shannon entropy) at time $t$, then the extension
of the second law of thermodynamics in NESS becomes
\begin{thm}
\begin{equation}\label{min-work}
\triangle S=\triangle \langle\phi\rangle\geq -\beta\langle Q_{ex}\rangle.
\end{equation}
Moreover, $-\beta\langle Q_{ex}\rangle=\triangle S$ if and only if
\begin{equation}
\int_0^T \frac{\partial \log\pi_s}{\partial
s}(X_s(\omega))ds=0,~P_{[0,T]}-a.s.\nonumber
\end{equation}
If the inhomogeneous multidimensional diffusion process
$X=\{X(t)\}_{0\leq t\leq T}$ is unperturbed, then $-\beta \langle
Q_{ex}\rangle=\triangle S$ for each trajectory $\omega$.
\end{thm}

Therefore, the generalized entropy difference $\triangle S$ between
two steady states can be measured through $-\beta \langle
Q_{ex}\rangle$ resulting from a slow (unperturbed) process
connecting these two states, which allows one to define the
generalized entropy of nonequilibrium steady states experimentally,
by measuring the excess heat obtained in a slow process between any
nonequilibrium steady state and an equilibrium state whose entropy
is known.

As in \cite{HS}, from (\ref{min-work}) one can derive the minimum work principle for steady-state thermodynamics.

\section*{Acknowledgement}

The authors would like to thank Prof. Hong Qian at University of
Washington for introducing reference \cite{HuS} and Prof. Min Qian
at Peking University for helpful discussion. This work is partly
supported by the NSFC 10701004.


\end{document}